\documentclass[%
showkeys,
preprint,
 amsmath,amssymb,
 aps,
]{revtex4-2}

\usepackage[hypertex]{hyperref}

\usepackage{graphicx}
\usepackage{dcolumn}
\usepackage{bm}

\newcommand{\tr}{\textrm{Tr}}

\newtheorem{defn}{Definition}
\newtheorem{remark}{Remark}

\newcommand{\ketbra}[1]{\left| {#1} \right\rangle}
\newcommand{\ketbraket}[1]{\left| {#1} \right\rangle \left\langle {#1}\right|}

\usepackage[T1]{fontenc}

\begin{document}


\title{Kolmogorov complexity of  unitary transformations\\ in quantum computing} 

\author{A. Kaltchenko}
\email{akaltchenko@wlu.ca}
\affiliation{Wilfrid Laurier University, Waterloo, Ontario, Canada}


\begin{abstract}
We introduce a notion of Kolmogorov complexity of unitary transformation, which can (roughly) be understood as the least possible amount of information required to fully describe and reconstruct a given finite unitary transformation. In the context of quantum computing, it corresponds to the least possible amount of data to define and describe a quantum circuit or quantum computer program.

Our Kolmogorov complexity of unitary transformation is built upon Kolmogorov "qubit complexity" of Berthiaume, W. Van Dam and S. Laplante  via mapping from unitary  transformations to positive operators, which are subsequently "purified". We discuss the optimality of our notion of Kolmogorov complexity in a broad sense and obtain a simple complexity bound.
\end{abstract}

\keywords{quantum information theory; information theory; quantum; Kolmogorov complexity; quantum computing; quantum information; unitary; quantum state.}
\maketitle
%

\section{Introduction}\label{Introduction}

In computer science \cite{li1993an} and information theory \cite{cover1997information}, Kolmogorov complexity of a (classical) string (for example, a binary sequence) or, more generally, of a classical finite object is commonly known as the shortest binary program~$p$, which runs on a Turing machine and computes (in other words, describes) the string or the object. Thus, Kolmogorov complexity is also called \emph{descriptive complexity}.

There have been several approaches~\cite{Vitanyi2001quantum,Gacs2001,Berthiaume2001quantum,Mora2005,Mora2007,Brudno} to define a complexity of a quantum state.
In~\cite{Vitanyi2001quantum,Berthiaume2001quantum,Mora2007}, the concept of Kolmogorov complexity was extended from classical to quantum domain. Such an extension naturally relies upon the concept of a universal quantum Turing machine, which was first proposed by Deutsch\cite{Deutsch}. The detailed construction of  quantum Turing machines can be found in~\cite{Vitanyi2001quantum,ADMDH,Vitanyi2001quantum}.

In~\cite{Berthiaume2001quantum}, the quantum Kolmogorov complexity, also known as "qubit complexity", is defined as follows:
the quantum Kolmogorov complexity of a string of qubits is defined, relative to a universal quantum Turing machine~$M$, as the length of the shortest qubit string which when given as input to~$M$, produces on its output register the qubit string. In this paper, we shall be concerned with the  "qubit complexity"\cite{Berthiaume2001quantum} type of quantum Kolmogorov complexity.

Let $\mathcal{H}$ be a 2-dimensional complex vector space (Hilbert space). A qubit is described by a unit vector in $\mathcal{H}$.  Then, for any integer~$n$, the state of~$n$ qubits corresponds to a unit vector in $n$-folded space $\mathcal{H}^{\otimes n}$. We will use the bra–ket  Dirac notation with low case Latin letters to denote vectors in~$\mathcal{H}^{\otimes n}$, that is $|x\rangle $, $|y\rangle $, etc.

Roughly speaking, for any~$\ketbra{x} \in \mathcal{H}^{\otimes n}$, the qubit complexity\cite{Berthiaume2001quantum} of ~$\ketbra{x}$ is equal to the logarithm in base~2 of the dimension of the smallest Hilbert space (spanned by computational basis vectors) containing a quantum state~$\ketbra{y}$ that, once fed into a universal quantum Turing machine, makes the universal quantum Turing machine compute the output~$\ketbra{x}$ and halt. Thus, qubit sequence~$\ketbra{y}$ can be seen as a {\em compressed description}  of~$\ketbra{x}$. The upper-bound for the qubit complexity of any~$\ketbra{x} \in \mathcal{H}^{\otimes n}$  is immediately seen to be~$n$.

We note that any unitary transformation~$U$ can be implemented\cite{nielsenquantum} as a quantum computation on universal quantum Turing machine (as well as using quantum logic circuits) and, vice versa, any quantum Turing machine computation (without measurements) can be seen as  a unitary transformation. Thus, to any unit vector~$\ketbra{x}$ in Hilbert space~$\mathcal{H}^{\otimes n}$, we assign its complexity equal to the quantity~$QC\left( \ketbra{x} \right)$, irrespective of the fact whether or not~$\ketbra{x}$ represents a  physical quantum system. The  complexity of~$\ketbra{x}$ shall be understood as the logarithm of the dimension of the respective Hilbert space.


Our notion of~$QC\left( \ketbra{x} \right)$  is compatible with the original qubit complexity\cite{Berthiaume2001quantum} as follows: If vector~$\ketbra{x}$ represents a quantum state, then the  quantity~$QC\left( \ketbra{x} \right)$  is also equal to the number of physical qubits as defined in~\cite{Berthiaume2001quantum}
and the quantum state can be fully described/represented by the number~$QC\left( \ketbra{x} \right)$ of qubits. {\em On the other hand, it's important to point out that if vector~$\ketbra{x}$ describes (i.e. maps) a unitary operator, it cannot be  described/represented by qubits}.

Our paper is organized as follows: In Section~\ref{chapter-unit-vector}, we define the Kolmogorov complexity of a unit vector in a Hilbert space, which  is based on   "qubit complexity"\cite{Berthiaume2001quantum} of Berthiaume, W. Van Dam and S. Laplante.  In Section~\ref{Chapter-Positive}, we describe the  "purification"\cite{nielsenquantum} procedure for quantum density operators and define the Kolmogorov complexity of a positive operator via  the  Kolmogorov complexity of the corresponding unit vector (i.e. purified state). In Section~\ref{chapter-unitary}, we introduce a mapping from a unitary  transformation to a positive operator. Then, we define the Kolmogorov complexity of the unitary  transformation via the Kolmogorov complexity of the constructed positive operator, discuss the operator convergence for approximate vector/operator, and obtain a simple complexity bound. In Section~\ref{chapter-Discussion}, we discuss the optimality of our notion of  Kolmogorov complexity  in the broadest  sense. In the Appendix, we summarize the properties of the quantum fidelity and its  relationship with the trace distance.

%
%
%

\section{Kolmogorov complexity of a unit vector}\label{chapter-unit-vector}

Berthiaume et. al in~\cite{Berthiaume2001quantum} defined "qubit complexity" and used the fidelity measure of how close two quantum states. We use their  definition of "qubit complexity" with the following two amendments:
\begin{enumerate}
  \item We restrict the original definition\cite{Berthiaume2001quantum} to pure quantum state states only (which corresponds to unit vectors).

  \item Instead of fidelity, we use the trace distance\footnote{Trace distance is also used in~\cite{Brudno}} as a measure  of how close two operators are.  As we summarize  in the Appendix, in view of relation~\eqref{Trace-equiv-fidelity}, the trace distance is essentially equivalent to the fidelity measure for quantum states. Moreover, the trace distance is well defined for all operators acting on Hilbert space, including  unitary transformations.
\end{enumerate}

The trace distance between operators~$\rho$ and~$\sigma$ is defined by
\begin{equation}
D(\rho,\sigma) \triangleq \frac{1}{2}\tr |\rho -\sigma|,
\end{equation}
where $|A| \triangleq \sqrt{A^\dag A}$. From this definition it follows that the trace distance is a genuine metric on quantum states, with $0 \leq D \leq 1$.
Note that in literature, trace distance is sometimes defined as $\tr |\rho -\sigma|$, without factor~$\frac{1}{2}$, which directly matches the operator trace norm $\|A\|_1 \equiv \tr \left(  \sqrt{A^\dag A} \right)$.

\begin{defn}: {\bf Kolmogorov complexity of a unit vector with trace distance $\delta$}:
For any quantum Turing machine M and pure qubit state~$\ketbra{x}$, the $\delta$-approximation Kolmogorov complexity, denoted~$QC^{\delta}_M(\ketbra{x})$, is the length of the shortest qubit string ~$\ketbra{y}$ such that, for any~$\delta >0$, we have $D\left(\ketbra{x}, M(\ketbra{y})  \right) \leq \delta$. We also introduce a notation
\begin{equation}\label{qubit-Compress}
Compress^{\delta}_M(\ketbra{x}) \triangleq  \ketbra{y};
\end{equation}
and a notation~$Reconstr^{\delta}_M(\ketbra{x})$ -- for vector~$ M\left( \ketbra{y} \right)$:
\begin{equation}\label{qubit-Reconstr}
Reconstr^{\delta}_M(\ketbra{x}) \triangleq  M\left( \ketbra{y} \right).
\end{equation}

\end{defn}
In the above definition, we slightly abuse notation by using vectors as the inputs for the trace distance $D(\cdot, \cdot)$, in which we understand any such  vector~$\ketbra{x}$ as operator~$\ketbraket{x}$.

\begin{remark}
Where the unit vector~$\ketbra{x}$ represents a quantum state of a physical system, the internal workings of the quantum Turing machine become relevant. Overwise, we are only concerned with the  input and output of the unitary transformation  implemented by QTM. The quantity $QC \left( \ketbra{x} \right)$ shall be understood as the logarithm of the dimension of the respective Hilbert space. The invariance of Kolmogorov complexity\cite{Berthiaume2001quantum}  and all the convergence results of~\cite{Berthiaume2001quantum}   remain applicable.
\end{remark}

If the trace distance~$\delta$ is  equal zero, we have the following definition.
\begin{defn}: {\bf Kolmogorov complexity of a unit vector with perfect reconstruction}: The perfect  reconstruction Kolmogorov complexity is~$QC^0_M(\ketbra{x})$.
\end{defn}
Note the notation difference: $QC^1_M(X)$ of~\cite{Berthiaume2001quantum} is equal to~$QC^0_M(\ketbra{x})$ of this paper, where the unit vector $\ketbra{x}$ is the state of qubit string $X$~in~\cite{Berthiaume2001quantum}.

\begin{remark}
Thought the paper, we will omit the subscript~M in the notation~$Compress^{\delta}_M(\cdot)$, $Reconstr^{\delta}_M(\cdot)$, $QC^{\delta}_M(\cdot)$, and other such complexity notations, which assume a quantum Turing machine~M. Additionally, we  will omit the superscript whenever $\delta =0$, that is for the case of perfect  reconstruction.
\end{remark}

\section{Kolmogorov complexity of a positive operator with trace~$1$}\label{Chapter-Positive}

Let~$\rho$ be a  positive  operator with trace~$1$.  
 We use a common notation $|i\rangle$ for the eigenvector basis of~$\rho$, thus,~$\rho$ has the following orthonormal decomposition:
\begin{equation}
\rho = \sum\limits_{i=1}^{2^n} p_i |i\rangle \langle i|,
\end{equation}
where~$\sum\limits_{i=1}^{2^n} p_i  =1$.

Now we are going to "purify"~$\rho$, using  "purification" procedure\cite{nielsenquantum}  as follows:

We note that operator~$\rho$ acts on the Hilbert space~$\mathcal{H}^{\otimes n}$ and we label the underlying system by~$A$. To purify~$\rho$, we introduce another, axillary system, which we label by~$R$ and which has a state space identical with~$A$, with orthonormal basis states~$|i^R\rangle$. We now define a  pure state for the combined system with the state space~$\mathcal{H}^{\otimes n} \otimes \mathcal{H}^{\otimes n}$ as follows:
\begin{equation}
|AR\rangle \triangleq \sum\limits_i \sqrt{p_i} |i^A\rangle |i^R\rangle.
\end{equation}

It's easy to see that the operator~$\rho$ can be obtained from the pure state~$|AR\rangle$ by tracing out the auxiliary system~$R$ as follows:
\begin{equation}\label{trace-notation}
\begin{split}
\tr_R\left( |AR\rangle \langle AR| \right)  & =  \sum\limits_{ij} \sqrt{p_ip_j} |i^A\rangle \langle j^A| \tr \left(  |i^R\rangle \langle j^R|\right) \\
& = \sum\limits_{ij} \sqrt{p_ip_j} |i^A\rangle \langle j^A | \delta_{ij}  \\
& = \sum\limits_i p_i |i^A\rangle \langle i^A | \\
& = \rho.\\
\end{split}
\end{equation}

To summarize the above construction,  for each~$\rho$, there is a corresponding unit vector~$|AR\rangle$:
\begin{equation}\label{mapping_U_AR}
\rho  \; \xrightarrow{\text{purifying} \ \rho } |AR\rangle
\end{equation}
Now, we are going to define the Kolmogorov complexity of Positive operator~$\rho$ with trace~$1$ as the complexity of  a "purified" unit vector~$|AR\rangle$. Note that~$\rho$ can be easily computed from~$|AR\rangle$ and vice versa, requiring only a finite, bounded amount of information about the auxiliary system used in purification. So for any meaningful definition of Kolmogorov complexity,   the complexity of $\rho$ will be equal to the complexity of~$|AR\rangle$ up to~$O(1)$. To diminish this~$O(1)$ to zero and mitigate the effect of purification on the optimality of our complexity definition, we will minimize over all purifications as follows:

\begin{defn}
\begin{equation}\label{def-of-K-Positive}
K^{\delta}(\rho)\triangleq \min_{|AR\rangle}  \{  QC^{\delta}\left( |AR\rangle \right) : \tr_R\left( |AR\rangle \langle AR| \right) = \rho \}.
\end{equation}
\end{defn}
Now we are going to show the convergence for approximate reconstruction. Let~$\ketbra{AR}$ be a purification which minimizes the right-hand side of~\eqref{def-of-K-Positive}. Let $\ketbra{\widetilde{AR}}$  be the a vector $Reconstr^{\delta}_M(\ketbra{AR})$ as defined in~\eqref{qubit-Reconstr}, so we have~$D\left( \left(\ketbraket{AR}\right), \left( \ketbraket{\widetilde{AR}}\right) \right) = \delta$.

The partial-trace operation  does not increase\cite{nielsenquantum} the trace distance, therefore, the following inequality holds:
\begin{equation}
D\left( \tr_R \left(\ketbraket{AR}\right), \tr_R \left( \ketbraket{\widetilde{AR}}\right)  \right) \leq D\left( \left(\ketbraket{AR}\right), \left( \ketbraket{\widetilde{AR}}\right) \right)
\end{equation}
Thus, we have:
\begin{equation}
D\left(  \rho, \tilde \rho \right) \leq D\left( \left(\ketbraket{AR}\right), \left( \ketbraket{\widetilde{AR}}\right) \right) = \delta,
\end{equation}
where~$\rho \equiv  \tr_R \left(\ketbraket{AR}\right)$ and $\tilde \rho  \equiv \tr_R \left( \ketbraket{\widetilde{AR}}\right)$.

Therefore, the convergence of the purification~$\ketbra{AR}$  implies the convergence of the positive operator~$\rho$. It makes the notion of Kolmogorov complexity $K^{\delta}(\rho)$ meaningful not only for perfect reconstruction (i.e. $\delta = 0$)  , but also for approximate one  (i.e. $\delta > 0$).

\begin{remark}
The generalization of Kolmogorov complexity from that for pure quantum states to that for  mixed states via purification was studied in~\cite{Mora2007}.
\end{remark}

\section{Kolmogorov complexity of  a unitary transformation}\label{chapter-unitary}
In this section, we provide a new, universal definition of the Kolmogorov complexity of a unitary transformation~$U$, via the Kolmogorov complexity of a (purified state) unit vector.

Let~$U$ be a unitary transformation on~$\mathcal{H}^{\otimes n}$, represented by a $2^n \times 2^n$ unitary matrix. 

First, we define an operator~$S$:
\begin{equation}
S \triangleq i\ln U.
\end{equation}
Since operator~$U$ is unitary, operator~$S$ is necessarily Hermitian. The logarithm of a unitary matrix is not uniquely defined as is the logarithm of a complex number. We can select~$S$ with eigenvalues in the range $0$  to $2\pi$ as follows. A unitary operator is a diagonalizable operator whose eigenvalues all have unit norm.  In  the eigenvector basis of~$U$,  we have a matrix of the following form:
\begin{equation}
  U =\begin{pmatrix}
    e^{-i\varphi_1} & 0 & \dots & 0 \\
    0 & e^{-i\varphi_2} & \dots & 0 \\
    \vdots & \vdots & \ddots & \vdots \\
    0 & 0 & \dots & e^{-i\varphi_{2^n}}
  \end{pmatrix},
\end{equation}
with the eigenvalues  of the form~$e^{-i\varphi_i}$. So the eigenvalues of ~$S$ will be
\begin{equation}
\varphi_i = i\ln U_{ii},
\end{equation}
where~$\varphi_i \in [0, 2\pi]$ and~$i =1,\ldots, 2^n $.

We note that~$S$ is a non-negative-definite operator with non-negative eigenvalues, however,  the trace~$\tr(S)$ is {\em not} necessarily equal to~1.  So we are going to purify the normalized version~$\frac{S}{\tr(S)}$. 

Finally, we define the Kolmogorov complexity of a unitary transformation~$U$ as the Kolmogorov  complexity of the normalized positive operator~$\frac{S}{\tr(S)}$:
\begin{defn}
\begin{equation}\label{def-of-K-U}
K^{\delta}(U)\triangleq K^{\delta} \left( \frac{S}{\tr(S)} \right).
\end{equation}
\end{defn}
To summarize the underlying construction,  for each unitary transformation~$U$ and $\delta \geq 0$, we construct a corresponding unit vector~$|AR\rangle$:
\begin{equation}\label{mapping_U_AR}
U \rightarrow \; S  \; \xrightarrow{\text{purifying} \ \frac{S}{\tr(S)} } |AR\rangle.
\end{equation}


Clearly, the upper bound on~$K^{\delta}(U)$ is equal to the logarithm in base~2 of the dimension of the Hilbert space~$\mathcal{H}^{\otimes n} \otimes \mathcal{H}^{\otimes n}$, i.e.~$2n$.

The exponential of a normal operator  is well defined and continuous operator function; and the partial-trace operation  does not increase\cite{nielsenquantum} the trace distance as discussed in Section~\ref{Chapter-Positive}. Therefore, the convergence of the $|AR \rangle$ reconstruction implies the convergence of~$e^{-i \tr(S) \cdot \tr_R\left( \ketbraket{AR} \right) }$  to~$U$ with respect to the operator trace norm~$\|A\|_1$, where~$\tr_R(\cdot)$ is the partial trace notation as in~\eqref{trace-notation}.

\begin{remark}
Essentially, unitary transformation~$U$  is mapped to a vector  $ \sqrt{\tr(i\ln U )} |AR \rangle$ of the lowest possible dimensionality. Given  vector~$\sqrt{\tr(i\ln U )} |AR \rangle$, we can reconstruct~$U$. Thus, vector~$\sqrt{\tr(i\ln U )} |AR \rangle$ can bees seen as the most compact description of~$U$.
\end{remark}

\section{Discussion}\label{chapter-Discussion}
In this section, we shall understand Kolmogov complexity in the most broad and permissive sense. We emphasise  that, despite the presence of word "complexity", the notion of  Kolmogov complexity is not about a computational time or a computational space complexity, but is about the shortest (most compact) possible description or representation of the object or system. In particular, the complexity of  one vector or  operator shall be equal (up to~$O(1)$) to the complexity of another vector or operator if one can be converted (mapped) to the other using a finite amount of information.


%

One needs an auxiliary classical or quantum Turing machine to compute matrix exponent and matrix logarithm as well as the  purification and partial trace. Logarithmic and exponential functions are known\cite{Muller} to have implementations  in Fortran and  C. The size of the program text file is 20--100~Kb. The  calculation precision can be arbitrary high and depends on  hardware, but does not affect the program size.  So are the Fortran and  C  programs for  other related calculations. Therefore, the Kolmogorov complexity of an auxiliary Turing machine configuration  to implement the mapping:
\begin{equation}\label{mapping_U_rho}
U \rightarrow  S
\end{equation}
is~$O(1)$. Similarly, as  discussed in Chapter~\ref{Chapter-Positive}, the Kolmogorov complexity of an auxiliary Turing machine configuration to implement the mapping:
\begin{equation}\label{mapping_rho_AR}
S \rightarrow |AR\rangle
\end{equation}
is~$O(1)$, too. 
\begin{remark}
 To obtain an infinitely high calculation precision we may need an auxiliary  Turing machine with an infinite tape. Nevertheless, the program size for the auxiliary calculations remains finite and does not depend on~$n$.
 \end{remark}

%

The~$O(1)$ complexity of mapping\eqref{mapping_U_rho}  and the~$O(1)$ complexity of  mapping\eqref{mapping_rho_AR}  imply that our complexity definitions of Positive operator and unitary transformation are optimal in the broad sense.

\section{Conclusion}
We have defined the Kolmogorov complexity of unitary transformations  and established its  relation with the qubit complexity. Thus, the concept  of Kolmogorov complexity has become ever more relevant  in quantum computing context.  It provides theoretical framework for quantum computing complexity analysis and can be further linked to the complexity of a quantum computer program at high level and low (hardware) level,  as well as quantum computer compilers.

\section{Appendix: The trace distance and Quantum fidelity}
In quantum computing, most commonly used are the following two measures of how close two quantum states are: quantum fidelity and the trace distance. Below we review their properties, for details, see~\cite{nielsenquantum}.

The trace distance: The trace distance between quantum states~$\rho$ and~$\sigma$ is defined by
\begin{equation}
D(\rho,\sigma) \triangleq \frac{1}{2}\tr |\rho -\sigma|,
\end{equation}
where $|A| \triangleq \sqrt{A^\dag A}$. From this definition it follows that the trace distance is a genuine metric on quantum states, with $0 \leq D \leq 1$. Note that in literature, trace distance is sometimes defined as $\tr |\rho -\sigma|$, without factor~$\frac{1}{2}$, which directly matches the operator trace norm $\|A\|_1 \equiv \tr \left(  \sqrt{A^\dag A} \right)$.

The fidelity between quantum states~$\rho$ and $\sigma$ is defined by
\begin{eqnarray}
F(\rho , \sigma) & \triangleq &
\tr\left({\sqrt{\sqrt{\rho}\cdot \sigma \cdot \sqrt{\rho}}}\right).
\end{eqnarray}
If~$\rho$ is a pure state~$\psi$, and~$\sigma$ is a pure state~$\varphi$, then the above definition reduces to~$|  \langle \psi   | \varphi \rangle |$. It can be shown that $0 \leq F(\rho , \sigma) \leq 1$. If $F(\rho , \sigma) = 1$, then $\rho=\sigma$, and vice versa.

For all purifications~$\psi$ of~$\rho$ and~$\varphi$ of~$\sigma$, the following two inequalities hold:
\begin{equation}
F(\rho , \sigma) \geq F \left(\ketbraket{\psi} , \ketbraket{\varphi}\right)  \equiv |  \langle \psi   | \varphi \rangle |
\end{equation}
\begin{equation}
D(\rho , \sigma)  \leq D \left(\ketbraket{\psi} , \ketbraket{\varphi}\right)
\end{equation}

Although not a metric, the fidelity  upper bounds and lower bounds the trace distance, so whenever the fidelity converges to one, the trace distance converges to zero:
\begin{equation}\label{Trace-equiv-fidelity}
1- F(\rho , \sigma) \leq D(\rho , \sigma) \leq \sqrt{1- F^2(\rho , \sigma)}
\end{equation}

\section{Data Availability Statement}
All data generated or analysed during this study are included in this published article.

\bibliography{myBibliography}

\end{document}